\documentclass[aps,prd,showpacs,twocolumn,showkeys,preprintnumbers]{revtex4}

\usepackage{graphicx}                                 
\usepackage{amsmath,amsfonts,mathbbol}

\newcommand{\ie}{\textit{\mbox{i.e.\ }}}              
\newcommand{\eg}{\textit{\mbox{e.g.\ }}}              
\renewcommand{\Re}{\mathfrak{Re}\,}                   
\newcommand{\Tr}{\mbox{Tr}}                           
\newcommand{\identity}{\mathbb{1}}                    
\newcommand{\bc}{\texttt{bc}}                         
\newcommand{\cbc}{\texttt{cbc}}                       
\newcommand{\fc}{\texttt{fc}}                         
\newcommand{\Ncp}{N_{\rm cp}{}}                       
\newcommand{\Fig}[1]{Fig.~\ref{#1}}
\newcommand{\Tab}[1]{Table~\ref{#1}}
\newcommand{\Sec}[1]{Sec.~\ref{#1}}
\newcommand{\Eq}[1]{Eq.~(\ref{#1})}

\begin{document}


\preprint{HU--EP--05/24} \preprint{LU-ITP 2005/008}
\title{Towards the infrared limit in $\boldsymbol{SU(3)}$ Landau gauge
       lattice gluodynamics} 
\author{A.~Sternbeck, E.--M.~Ilgenfritz, M.~M\"uller--Preussker}
\affiliation{Humboldt-Universit\"at zu Berlin, Institut f\"ur Physik,
  D-12489 Berlin, Germany}
\author{A.~Schiller}
\affiliation{Universit\"at Leipzig, Institut f\"ur Theoretische
  Physik, D-04109 Leipzig, Germany}

\date{July 12, 2005}

\begin{abstract}
  We study the behavior of the gluon and ghost dressing functions in
  $SU(3)$ Landau gauge at low momenta available on lattice sizes $12^4-32^4$ at
  $\beta=5.8$, $6.0$ and $6.2$. We demonstrate the ghost dressing function
  to be systematically dependent on the choice of Gribov copies,
  while the influence on the gluon dressing function is not resolvable.
  The running coupling given in terms of these functions
  is found to be decreasing for momenta $q<0.6$~GeV. 
  We study also effects of the finite volume and of the lattice discretization.
\end{abstract}


\maketitle

\section{Introduction}
\label{sect:intro}

Studying the relation of non-perturbative features of QCD, such as
confinement and dynamical chiral symmetry breaking, to the properties
of propagators, there are two popular approaches at present: lattice
gauge theory and Dyson-Schwinger equations (DSE). The latter approach
allows to address directly the low-momentum region for a coupled
system of quark, gluon and ghost propagators which is of interest
for hadron physics \cite{Alkofer:2000wg}. In particular their infrared
behavior could be related to the mechanism of dynamical symmetry breaking
and to confined gluons \cite{Alkofer:2000wg,Fischer:2003rp}.

In fact, the DSE approach has revealed that in the infrared momentum
region a diverging ghost propagator $G$ is intimately connected with a
suppressed gluon propagator $D_{\mu\nu}$. In Landau gauge, they can be
written as
\begin{eqnarray}
  \label{eq:Zgl}
  D_{\mu\nu}(q) & = & \left(\delta_{\mu\nu} -
    \frac{q_{\mu}~q_{\nu}}{q^2} \right) \frac{Z_{D}(q^2)}{q^2} \; ,\\
  \label{eq:Zgh}
  G(q) & = & \frac{Z_{G}(q^2)}{q^2} \; .
\end{eqnarray}
Here $Z_{D}(q^2)$ and $Z_{G}(q^2)$ denote the dressing functions of
the corresponding propagators. They describe the deviation
from the momentum dependence of the free propagators.
Based on the Dyson-Schwinger approach and under mild assumptions these
functions are predicted to behave in the limit $q^2 \to 0$ as
follows \cite{Alkofer:2000wg}:
\begin{equation}
  Z_{D}(q^2) \propto (q^2)^{\kappa_D} \; , \quad Z_{G}(q^2) \propto
  (q^2)^{-\kappa_G} \;
  \label{eq:infrared-behavior}
\end{equation}
with exponents satisfying $\kappa_D = 2\kappa_G$. In Landau gauge
$\kappa_G \approx 0.595$ \cite{Lerche:2002ep,Zwanziger:2001kw}.
Thus the ghost propagator diverges stronger than $1/q^2$ and the gluon
propagator is infrared suppressed. This is in agreement with the
Zwanziger-Gribov horizon condition
\cite{Zwanziger:2003cf,Zwanziger:1993dh,Gribov:1977wm} as well as with
the Kugo-Ojima confinement criterion
\cite{Kugo:1979gm}. Zwanziger~\cite{Zwanziger:2003cf} has suggested
that in the continuum this behavior of the propagators in Landau gauge
results from the restriction of the gauge fields to the Gribov region
$\Omega$, where the Faddeev-Popov operator is non-negative.

Using further the ghost-ghost-gluon vertex, the gluon and ghost dressing
functions can be used to determine a renormalization group invariant
running coupling in a momentum subtraction scheme
as \cite{vonSmekal:1997is,vonSmekal:1998,Bloch:2003sk}
\begin{equation}
  \label{eq:runcoupling}
  \alpha_s(q^2) = \frac{g^2_0}{4\pi}\, Z^2_{G}(q^2)\,Z_{D}(q^2)
\end{equation}
which then enters the quark DSE \cite{Alkofer:2000wg,Fischer:2003rp}.
This definition relies on the fact that the ghost-ghost-gluon vertex
renormalization function $Z_1(q^2)$ is constant, which is true at
least to all orders in perturbation theory
\cite{Taylor:1971ff}. Indeed, a recent numerical investigation of
$Z_1$ for the $SU(2)$ case shows that also nonperturbatively
$Z_1$ is finite and constant \cite{Cucchieri:2004sq}. Applying the
behavior given in \Eq{eq:infrared-behavior} the running
coupling approaches a finite value $\alpha_s(0) \approx 8.915/N$
for $SU(N)$ at zero momentum in the DSE approach \cite{Lerche:2002ep}.

Nevertheless, numerical investigations of those features in lattice
simulations are still necessary to check to what extent the truncation
of the coupled set of DSEs influences the final result. There are
several studies in Landau gauge which confirm the anticipated behavior
at least for the $SU(2)$ case
\cite{Bloch:2003sk,Langfeld:2001cz,Gattnar:2004bf,Bloch:2002we}. Also
lattice studies (\cite{Oliveira:2004gy,Sternbeck:2004xr,Sternbeck:2004qk},
\cite{Furui:2004cx,Furui:2003jr} and references therein) for the
$SU(3)$ case indicate the correctness of the proposed infrared
behavior. However, as recent DSE investigations show
\cite{Fischer:2005,Fischer:2002eq,Fischer:2002hn} the infrared
behavior of the gluon and ghost dressing functions and of the running
coupling is changed on a torus. In particular, the running coupling
decreases at low momenta.

This paper presents a lattice study of the gluon and ghost
dressing functions and of the running coupling at low momenta in $SU(3)$
Landau gauge. We also
focus, more carefully than usually, on the problem of the Gribov
ambiguity in lattice simulations. In continuum, a gauge orbit has more
than one intersection (Gribov copies \cite{Gribov:1977wm}) with the
transversality plane (where $\partial_{\mu} A_{\mu} = 0$
holds for the gauge potential $A_{\mu}$) within the
Gribov region $\Omega$. Expectation values taken over this
region are argued to be equal to those over the fundamental modular
region $\Lambda$ which includes only the absolute maximum of the gauge
functional~\cite{Zwanziger:2003cf}.

On a finite lattice, however, this
equality cannot be expected~\cite{Zwanziger:2003cf}. In the
literature it is widely taken
for granted that the gluon propagator does not depend on the choice of
Gribov copy, while an impact on the $SU(2)$ ghost propagator has been
observed \cite{Cucchieri:1997dx,Bakeev:2003rr,Nakajima:2003my}.
However, in a more recent investigation~\cite{Silva:2004bv} an
influence of the Gribov copies ambiguity on the $SU(3)$ gluon
propagator has been demonstrated, too. Here we assess the
importance of the Gribov ambiguity on a finite lattice for the $SU(3)$
ghost propagators.

This paper is structured as follows: In \Sec{sect:definitions} we shall
define all quantities which are investigated in this study. Then,
after specifying the lattice setup used, the dependence of the gluon
and ghost propagator on the choice of Gribov copies 
and lattice discretization as well as finite-volume effects are 
discussed in \Sec{sect:propagators}.  
We also discuss the problem of exceptional gauge copies in
\Sec{sect:exceptional}. In \Sec{sect:running_coupling} the behavior of the
running coupling is presented. In Appendix \ref{sect:pcg} we show
how the inversion of the F-P operator can be accelerated by
pre-conditioning with a Laplacian operator.

\section{Definitions}
\label{sect:definitions}

To study the ghost and gluon propagators using lattice simulations
one has to fix the gauge for each thermalized $SU(3)$ gauge field
configuration $U=\{U_{x,\mu}\}$. We adopted the Landau gauge condition which
can be implemented by searching for a gauge transformation
\begin{displaymath}
  U_{x,\mu} \rightarrow {}^{g}U_{x,\mu}=g_x\,
  U_{x,\mu}\,g^{\dagger}_{x+\hat{\mu}}
\end{displaymath}
which maximizes the Landau gauge functional
\begin{equation}
  F_{U}[g] = \frac{1}{4V}\sum_{x}\sum_{\mu=1}^{4}\Re\Tr \;{}^{g}U_{x,\mu}
  \label{eq:functional}
\end{equation}
while keeping the Monte Carlo configuration $U$ fixed.
Here $g_x$ are elements of $SU(3)$.

The functional $F_{U}[g]$ has many different local maxima which can
be reached by inequivalent gauge transformations $g$, the number of
which increases with the lattice size. As the inverse coupling constant
$\beta$ is decreased, increasingly more of those maxima become accessible
by an iterative gauge fixing process starting from a given (random)
gauge transformation $g_x$. The different gauge copies corresponding to
those maxima are called \emph{Gribov copies}, due to their resemblance to
the Gribov ambiguity in the continuum \cite{Gribov:1977wm}. All Gribov
copies $\{{}^{g}U\}$ belong to the same gauge orbit created by the Monte
Carlo configuration $U$ and satisfy the differential Landau gauge
condition (lattice transversality condition)
$(\partial_{\mu}{}^{g}\!\!A_{\mu})(x) = 0$ where
\begin{equation}
  (\partial_{\mu}{}^{g}\!\!A_{\mu})(x) \equiv {}^{g}\!\!A_{\mu}( x+
  \hat\mu/2)-{}^{g}\!\!A_{\mu}( x- \hat\mu/2).
  \label{eq:transcondition}
\end{equation}
Here ${}^{g}\!\!A_\mu(x+\hat{\mu}/2)$ is
the non-Abelian (hermitian) lattice gauge potential which may be
defined at the midpoint of a link
\begin{eqnarray}\nonumber
  {}^{g}\!\!A_\mu(x+\hat{\mu}/2) &=&\frac{1}{2iag_0}\left(^{g} U_{x,\mu} -
    \ ^{g} U^{\dagger}_{x,\mu}\right)\\
   &&-\frac{\identity}{6iag_0}\Tr\left(^{g} U_{x,\mu} -
    \ ^{g} U^{\dagger}_{x,\mu}\right)\: .
\label{eq:A-definition}
\end{eqnarray}
In this way it is accurate to $\mathcal{O}(a^2)$. The bare gauge
coupling $g_0$ is related to the inverse lattice coupling via
$\beta=6/g^2_0$ in the case of $SU(3)$.
In the following, we will drop the label $g$ for convenience, \ie we
consider $U$ to be already put into the Landau gauge such that
$g=\identity$ maximizes the functional in \Eq{eq:functional} relative
to the neighborhood of the identity.

The gluon propagator $D^{ab}_{\mu\nu}(q^2)$ is the Fourier transform
of the gluon two-point function, {\it i.e.} the expectation
value
\begin{equation}
  D^{ab}_{\mu\nu}(q) = \left\langle \widetilde{A}^a_{\mu}(k)
  \widetilde{A}^b_{\nu}(-k) \right\rangle = \delta^{ab} D_{\mu\nu}(q)
\label{eq:D-definition}
\end{equation}
which is required to be color-diagonal. Here $\widetilde{A}^a_{\mu}(k)$
is the Fourier transform of $A^a_\mu(x+\hat{\mu}/2)$ and $q$ denotes
the momentum
\begin{equation}
  q_{\mu}(k_{\mu}) = \frac{2}{a} \sin\left(\frac{\pi
      k_{\mu}}{L_{\mu}}\right)
\label{eq:p-definition}
\end{equation}
which corresponds to a integer-valued lattice momentum $k$.
Since $k_{\mu} \in \left(-L_{\mu}/2, L_{\mu}/2\right]$
the lattice equivalent of $q^2(k)$ can be realized by different
$k$. According to Ref.~\cite{Leinweber:1998uu}, however, a subset of
lattice momenta $k$ has been considered only for the final analysis of
the gluon propagator, although the FFT algorithm provides us with all
lattice momenta. Details are given below.

Assuming reality and rotational invariance we envisage for the
(continuum) gluon propagator the general tensor structure:
\begin{equation}
  D_{\mu\nu}(q) = \left( \delta_{\mu\nu} - \frac{q_{\mu}~q_{\nu}}{q^2}
  \right) D(q^2) + \frac{q_{\mu}~q_{\nu}}{q^2}
        \frac{F(q^2)}{q^2}
\label{eq:decomposition}
\end{equation}
with $D(q^2)$ and $F(q^2)$ being scalar functions.
On the lattice these functions are extracted by projection
and are expected to scatter, rather than being smooth functions of
$q^2$. Using the Landau gauge condition the longitudinal form factor $F(p^2)$
vanishes. Recalling the mentioned Gribov ambiguity of the chosen gauge
copy there is no {\it a priori} reason to assume the estimator of
$D(q^2)$ is not influenced by the choice.

\begin{table*}
  \centering
  \begin{tabular}{r@{\qquad}c@{\quad}r@{\qquad}r@{\quad}r@{\qquad}l}
\hline\hline
    No. & $\beta$ & lattice & \# conf & \# copies &
    selected $k$ for $G(k)$\\
\hline
    1 & 5.8  & $24^4$  & 40 & 30 & $([1,0],0,0)$  \\
    2 & 6.2  & $12^4$  & 150 & 20 & $([1,0],0,0)$\\
    3 & 6.2  & $16^4$  & 100 & 30 & $([1,0],0,0)$ \\
    4 & 6.2  & $24^4$  & 35 & 30 & $([1,0],0,0)$\\*[0.1cm]
    5 & 5.8  & $16^4$  & 40 & 30 &  $\{(k,k,k,k), k=1\ldots6\}$,
         (2,1,1,1)\\
    6 & 6.0  & $16^4$  & 40 & 30 & $\{(k,k,k,k), k=1\ldots6\}$,
         (2,1,1,1)\\
    7 & 6.2  & $16^4$  & 40 & 30 & $\{(k,k,k,k), k=1\ldots6\}$,
         (2,1,1,1)\\*[0.1cm]
    8 & 5.8  & $24^4$  & 25 & 40 & $\{(k,k,k,k), k=1\ldots6\}$,
         (2,1,1,1), ([1,0,0,0]), (1,1,0,0), (1,1,1,0)\\
    9 & 6.0   & $24^4$  & 30 & 40 & $\{(k,k,k,k), k=1\ldots6\}$,
         (2,1,1,1), ([1,0,0,0]), (1,1,0,0), (1,1,1,0)\\
   10 & 6.2   & $24^4$  & 30 & 40 & $\{(k,k,k,k), k=1\ldots6\}$,
         (2,1,1,1), ([1,0,0,0]), (1,1,0,0), (1,1,1,0) \\*[0.1cm]
   11  & 5.8  & $32^4$  & 14 & 10 & (1,0,0,0), (1,1,0,0),
    (1,1,1,0), (1,1,1,1), (2,1,1,1) \\
\hline\hline
  \end{tabular}
\caption{Statistics of the data used in our final analysis. The 6th column
  lists all tuples of momentum $k$ the ghost propagator has been
  calculated for. If entries are given in squared bracket all of their
  permutations are meant.}
  \label{tab:statistic}
\end{table*}

The ghost propagator is derived from the Faddeev-Popov (F-P) operator,
the Hessian of the gauge functional given in \Eq{eq:functional}. We
expect that the properties of the F-P operator differ for the
different maxima of the functional (Gribov copies). This should have
consequences for the ghost propagator as is shown below.

After some algebra the F-P operator can be written in terms of the
(gauge-fixed) link variables $U_{x,\mu}$ as
\begin{eqnarray}
  M^{ab}_{xy} & = & \sum_{\mu} A^{ab}_{x,\mu}\,\delta_{x,y}
  - B^{ab}_{x,\mu}\,\delta_{x+\hat{\mu},y}
  - C^{ab}_{x,\mu}\,\delta_{x-\hat{\mu},y}\quad
  \label{eq:FPoperator}
\end{eqnarray}
with
\begin{eqnarray*}
  A^{ab}_{x,\mu} &=& \phantom{2\cdot\ } \Re\Tr\left[
    \{T^a,T^b\}(U_{x,\mu}+U_{x-\hat{\mu},\mu}) \right],\\
  B^{ab}_{x,\mu} &=& 2\cdot\Re\Tr\left[ T^bT^a\, U_{x,\mu}\right],\\
  C^{ab}_{x,\mu} &=& 2\cdot\Re\Tr\left[ T^aT^b\, U_{x-\hat{\mu},\mu}\right]\;.
\end{eqnarray*}
Here $T^a$ and $T^b$ are the (hermitian) generators of the
$\mathfrak{su}(3)$ Lie algebra satisfying $\Tr[T^aT^b]=\delta^{ab}/2$.

The ghost propagator is calculated as the following ensemble average
\begin{equation}
  G^{ab}(q) = \frac{1}{V} \sum_{x,y} \left\langle {\rm e}^{-2\pi i\,k
      \cdot (x-y)} [M^{-1}]^{ab}_{xy} \right\rangle_U\,.
  \label{eq:Def-ghost}
\end{equation}
It is diagonal in color space: $G^{ab}(q) = \delta^{ab} G(q)$.
Following Ref.~\cite{Suman:1995zg, Cucchieri:1997dx} we have used the conjugate
gradient (CG) algorithm to invert $M$ on a plane wave $\vec{
\psi}_c$ with color and position components \mbox{$\psi^a_c(x) = \delta^{ac}
\exp (2\pi i\,k\!\cdot\! x)$}. In fact, we applied the pre-conditioned
CG algorithm (PCG)
to solve $M^{ab}_{xy}\phi^{b}(y)=\psi^a_c(x)$. As pre-conditioning matrix
we used the inverse Laplacian operator $\Delta^{-1}$ with diagonal
color substructure. This significantly reduces the amount of computing
time as it is discussed in more detail in Appendix \ref{sect:pcg}.

After solving $M\vec{\phi}=\vec{\psi}_c$ the resulting
vector $\vec{\phi}$ is projected back on $\vec{\psi}_c$
such that the average $G^{cc}(q)$ over the color index $c$
can be taken explicitly. Since the F-P operator $M$ is singular if acting on
constant modes, only $k \ne (0,0,0,0)$ is permitted. Due to high
computational requirements to invert the F-P operator for each $k$,
separately, the estimator on a single, gauge-fixed configuration is
evaluated only for a preselected set of momenta $k$. In \Tab{tab:statistic} a
detailed list is given.

\begin{figure*}
\mbox{\includegraphics[height=7cm]{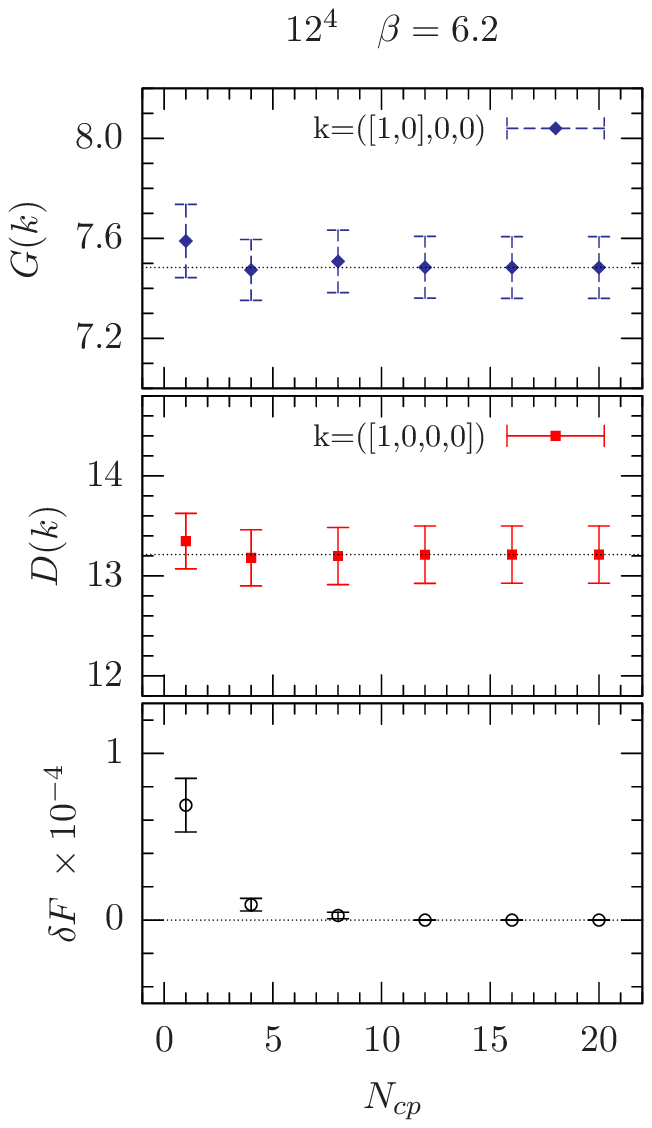}\quad
  \includegraphics[height=7cm]{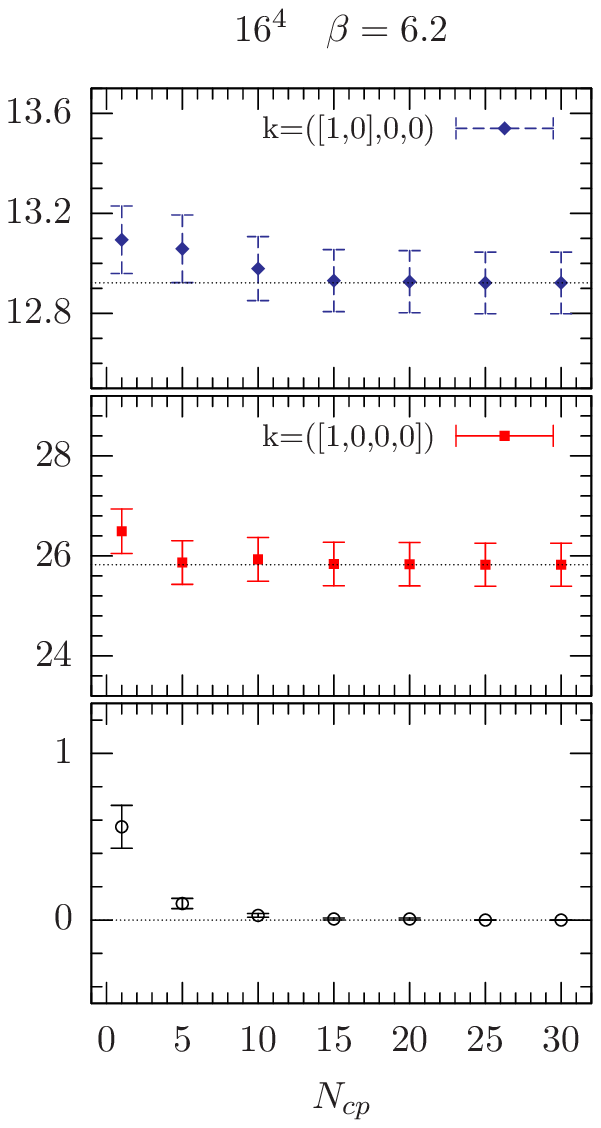}\quad
  \includegraphics[height=7cm]{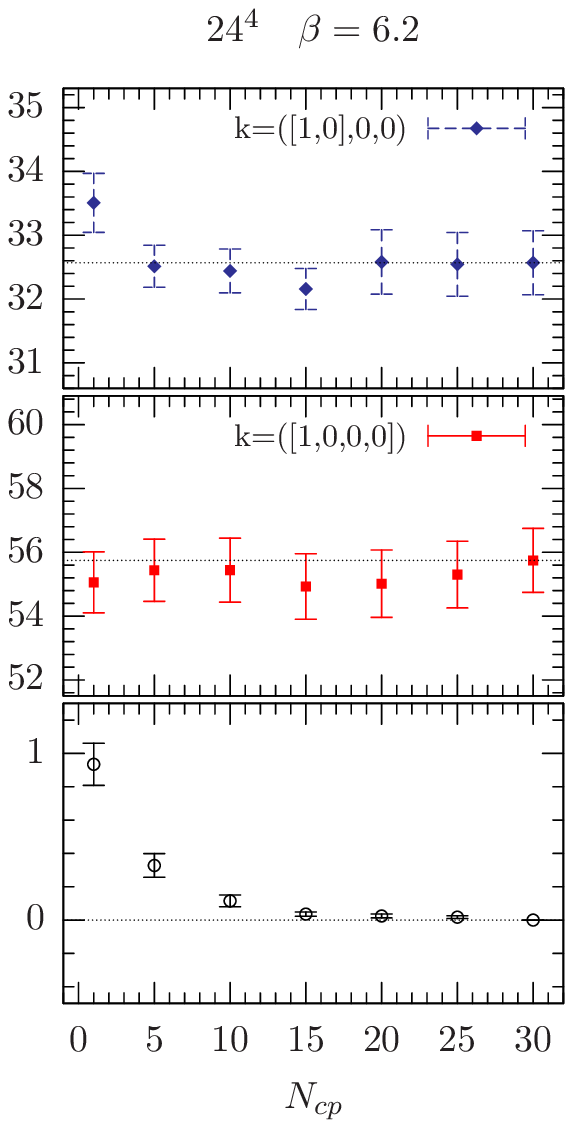}\quad
  \includegraphics[height=7cm]{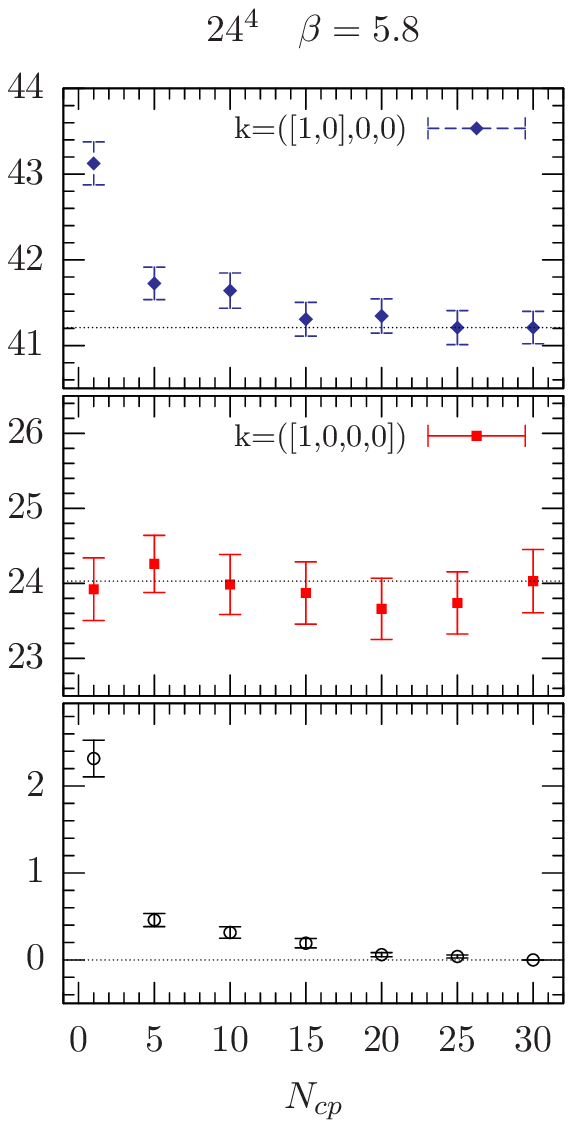}}
\caption{The upper panels show the ghost propagator $G(k)$ as average
 over two realizations $k=(1,0,0,0)$ and $k=(0,1,0,0)$ of the
 smallest lattice momentum, measured always on the \emph{best} gauge
 copy among $N_{\rm cp}$ copies. In the middle panels the same
 dependence is shown for the gluon $D(k)$ propagator,
 however, as average over all four permutations of $k=(1,0,0,0)$.
 The lower panels show the relative difference $\delta F =
 1-F^{\cbc}/F^{\bc}$ of the corresponding current best functional
 values $F^{\cbc}$ to the value $F^{\bc}$ of the overall best copy.}
\label{fig:gh_vs_cp}
\end{figure*}

\section{Results for the ghost and gluon propagators}
\label{sect:propagators}

\subsection{Lattice samples}

For the purpose of this study we have analyzed pure $SU(3)$ gauge
configurations which have been thermalized with the standard Wilson
action at three values of the inverse coupling constant $\beta=5.8$,
$6.0$ and $6.2$. For thermalization an update cycle of one
heatbath and four micro-canonical over-relaxation steps was used.
As lattice sizes we used $16^4$, $24^4$ and $32^4$.
For tests, exposing the inherent problems of the Gribov problem
under the aspect of volume dependence, also smaller lattices ($8^4$
and $12^4$) have been considered at lower cost.

To each thermalized configuration $U$ a random set of $\Ncp$
local gauge transformation $\{g_x\}$ were assigned. Each of those served
as starting point for a gauge fixing procedure for which we used
standard \emph{over-relaxation} with over-relaxation parameter tuned
to $\omega=1.63$. Keeping all $U_{x,\mu}$ fixed this iterative
procedure creates a
sequence of local gauge transformations $g_x$ at sites $x$ with
increasing values of the gauge functional (\Eq{eq:functional}).
Thus, the final Landau gauge is iteratively approximated until the
stopping criterion in terms of the transversality (see
\Eq{eq:transcondition})
\begin{equation}
  \max_x \left[\partial_{\mu} {}^g\!\! A_{\mu}(x)\right]^2 < 10^{-14}
  \label{eq:stop_crit}
\end{equation}
is fulfilled. Consequently, each random start results in its own local
maximum of the gauge functional. Certain
extrema of the functional are found multiple times. In fact, this
happened frequently on the small lattices, $8^4$ and $12^4$, but
rather seldom on larger lattices. Note that we used the maximum in relation
(\ref{eq:stop_crit}) which is very conservative. However,
the precision of transversality dictates how symmetric the
F-P operator $M$ can be considered. This is crucial for its inversion and
thus dictates the final precision of the ghost propagator.

To study the dependence on Gribov copies of the propagators, in the
course of $\Ncp$ repetitions for each $U$, the gauge copy with the
largest functional value is stored under the name \emph{best copy}
(\bc). The first gauge copy is also stored, labeled as \emph{first
copy} (\fc). However, it is as good as any other arbitrarily selected
gauge copy.

The more gauge copies one gets to inspect, the bigger the
likeliness that the copy labeled as \bc{} actually represents the
absolute maximum of the functional in \Eq{eq:functional}. With
increasing number $\Ncp$ of copies the expectation value of
gauge variant quantities, evaluated on \bc{} representatives, is
converging more or less rapidly as we will discuss next.

\subsection{How severe is the lattice Gribov problem for the propagators?}
\label{sec:gribov_problem}

First we present results of a combined study of the gluon and ghost
propagators on the same sets of \fc{} and \bc{} representatives of our
thermalized gauge field configurations. This allows us to assess the
importance of the Gribov copy problem.

Numerically, it turns out that the dependence of the ghost propagator on
the choice of the best copy is most severe for the smallest
momentum. In addition, this depends on the lattice size and
$\beta$. Therefore we studied first the dependence of the ghost
and gluon propagators at lowest momentum on the (same) best copies
as function of the number of gauge copies $N_{\rm cp}$ under
inspection. This was done at $\beta=6.2$ where we used $12^4$,
$16^4$ and $24^4$ lattices. The number of thermalized
configurations used for these three lattice sizes are given in
\Tab{tab:statistic}. To check the dependence on $\beta$ also a
simulation at $\beta=5.8$ on a $24^4$ lattice was performed.

\begin{figure*}
 \mbox{\includegraphics[width=8cm]{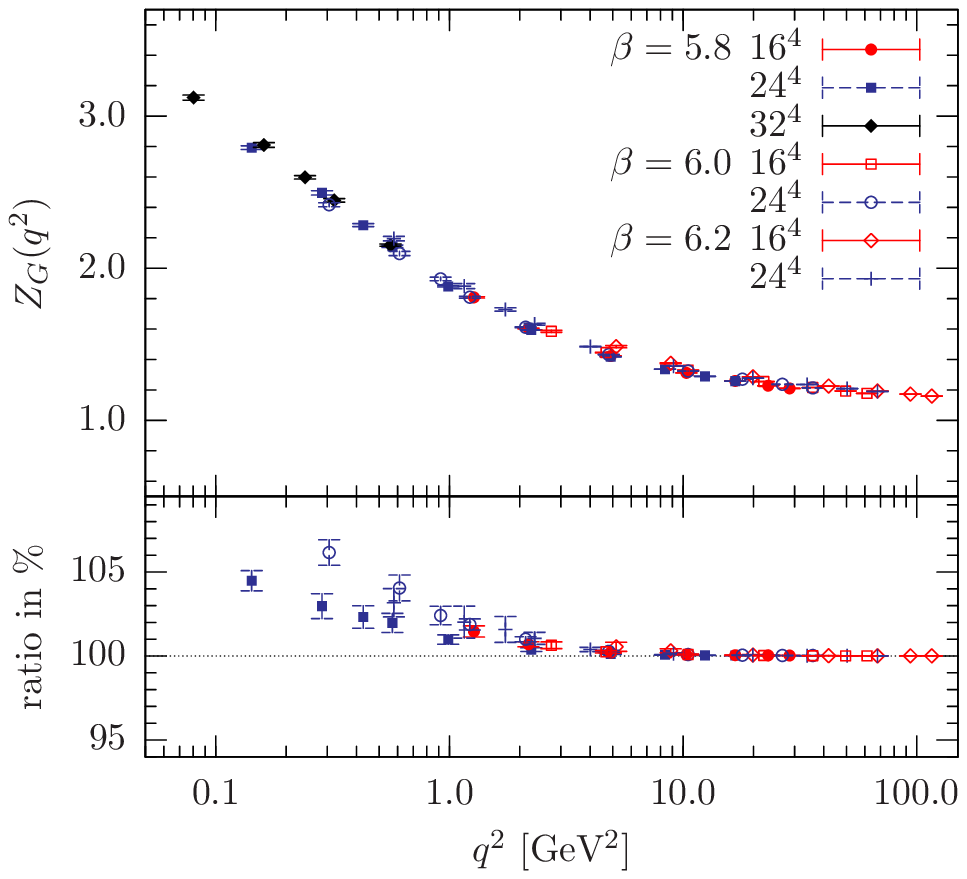}\quad
    \includegraphics[width=8cm]{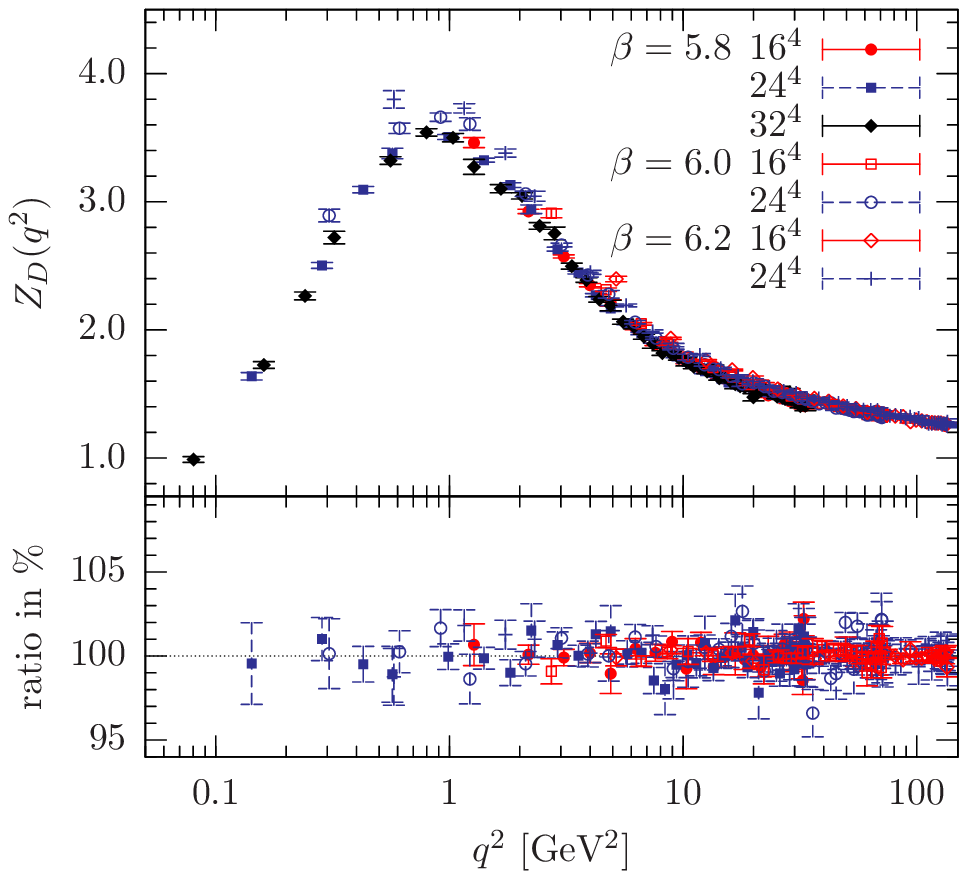}}
  \caption{The upper panels show the dressing functions of the
           ghost $Z_{G}$ and gluon propagator $Z_{D}$ measured on
           \emph{best} gauge copies as functions of the momentum $q^2$ (scaled
           to physical units at $\beta=5.8, 6.0$ and $6.2$) using
           various lattice sizes. The lower panels show the ratio
           $\langle Z^{\rm fc}\rangle / \langle Z^{\rm bc}\rangle$
           determined on first (\fc{}) and best (\bc{}) gauge copies
           using jackknife.}
     \label{fig:gh_and_gl_as_func_of_q}
\end{figure*}

The results of this investigation are show in
\Fig{fig:gh_vs_cp}. While there the ghost
propagator is shown as an average over the two realizations
\mbox{$k=(1,0,0,0)$} and \mbox{$k=(0,1,0,0)$} of the smallest lattice
momentum $q(k)$, the gluon propagator
has been averaged over all four non-equivalent realizations. Note
that $D(k)=D(-k)$. It is clearly visible that the expectation value of
the gluon propagator does not change within errors as $\Ncp$
increases, independent of the lattice size and $\beta$.
Contrarily, the ghost propagator at $\beta=5.8$ on a $24^4$ lattice
saturates (on average) if calculated on the best among $\Ncp=15$ gauge
copies. At $\beta=6.2$ the number of gauge fixings attempts reduces to
$5\le\Ncp\le 10$ on a $16^4$ and $24^4$ lattice. On the $12^4$ lattice a
small impact of Gribov copies is visible, namely $1<\Ncp\le5$.
The lower panels of \Fig{fig:gh_vs_cp} show the relative difference
$\delta F = 1-F^{\cbc}/F^{\bc}$ of the corresponding (current best)
functional values $F^{\cbc}$ to the value $F^{\bc}$ of the overall
best copy after $\Ncp=20$, respectively $\Ncp=30$, attempts. This
may serve as an indicator how
large $\Ncp$ has to be on average for the chosen algorithm to have found a
maximum of $F$ close to the global one.

In order to study further the low-momentum dependence of the gluon and
ghost dressing functions, $Z_{D}$ and $Z_{G}$, as given by
\Eq{eq:Zgl} and \Eq{eq:Zgh} we have performed similar simulations using
lattice sizes $12^4$, $16^4$, $24^4$ and $32^4$ at $\beta=5.8$, 6.0 and
6.2. Following Ref.~\cite{Necco:2001xg} these values of $\beta$
correspond to $a^{-1}$=1.446~GeV, 2.118~GeV and 2.914~GeV using the
Sommer scale $r_0=0.5$~fm. These values of the lattice spacing $a$
associated to $\beta$ turn out to be more appropriate as those
formerly used by us and others (see
\cite{Sternbeck:2004xr,Sternbeck:2004qk,Silva:2004bv,Leinweber:1998uu}).

We have fixed a conservative number of $\Ncp=30$ gauge copies
per thermalized configuration on a $16^4$ lattice and $\Ncp=40$
on a $24^4$ lattice. Both the gluon and the ghost propagator,
respectively their dressing functions, have been measured on the same set of
\fc{} and \bc{} copies. Due to the large amount of computing time
necessary for the $32^4$ lattice we could afford to measure the ghost
propagator for the first and best among only $\Ncp=10$ copies, which is
certainly not enough.

The data for the gluon propagator $D(k)$ have been determined for all momenta
at once. However, we used only a subset of momenta for the final
analysis. In fact, inspired by Ref.~\cite{Leinweber:1998uu}, only data
$D(k)$ with $k$ lying in a cylinder with radius of one momentum unit along
one of the diagonals $\hat{n}=1/2(\pm1,\pm1,\pm1,\pm1)$ have been selected.
Since we are using a symmetric lattice structure only data with $k$ satisfying
$\sum_{\mu}k^2_{\mu} - (\sum_{\mu}k_{\mu}\hat{n}_{\mu})^2\le 1$ are
surviving this cylindrical cut. In agreement with
\cite{Leinweber:1998uu} this recipe has drastically reduced lattice artifacts,
in particular for large momenta. Additionally, we try to keep finite
volume artifacts at lower momenta under control by removing all data
$D(k)$ with one or more vanishing momentum components $k_{\mu}$
\cite{Leinweber:1998uu}. However, this we have done only for data on a
$12^4$ and $16^4$ lattice. In \Sec{sec:sytematic} we shall discuss in
more detail finite volume effects at various momenta.

In view of this we have chosen appropriate sets of momenta for
the ghost propagator, as listed in \Tab{tab:statistic} in detail.

The final results of the dressing functions $Z_{D}$ and $Z_{G}$
measured on \bc{} copies are shown in the upper panels of
\Fig{fig:gh_and_gl_as_func_of_q}. All momenta $q(k)$ have been mapped
to physical momenta using the lattice spacings $a$ given above. As
expected the ghost dressing function diverges with decreasing
momenta, while the gluon dressing function decreases after passing a
turnover at about $q^2=0.8$~GeV$^2$. However, for the purpose of the
expected infrared behavior given in \Eq{eq:infrared-behavior} the
data for momenta \mbox{$q^2<0.25$~GeV$^2$} are not sufficiently
abundant to extract a critical exponent $\kappa_G > 0.5$ as expected
from the Dyson-Schwinger approach.  In particular, the fit parameters
are not stable under a change of the upper momentum cutoff. The best
fits give $\kappa_G \approx 0.25$.

In the lower panels of \Fig{fig:gh_and_gl_as_func_of_q} we present
the ratio of the dressing functions, $\langle Z^{\fc}\rangle /
\langle Z^{\bc}\rangle$, calculated using jackknife from first and
best gauge copies as a function of the momentum. 
There the data from simulations on a $32^4$ lattice have been 
excluded, since only $\Ncp=10$ gauge copies have been inspected which
would result in an underestimate of the ratio $\langle Z^{\fc}\rangle
/ \langle Z^{\bc}\rangle$. As is clear from
these panels we do not observe a systematic dependence on the choice
of Gribov copies for the gluon propagator. In contrast the ghost propagator is
systematically overestimated for \fc{} (arbitrary) gauge copies.
This effect holds up to momenta of about 2~GeV$^2$. 

Comparing also the ratios for the ghost propagator at
$q<1$~GeV, the rise at $\beta=6.0$ is obviously larger than that
at $\beta=5.8$. In both cases the data are from simulations on a
$24^4$ lattice. Thus, it seems that by increasing
the physical volume (lower~$\beta$) the effect of the Gribov ambiguity
gets smaller if the same physical momentum is considered. 
This cannot be due to a too small number $\Ncp$ of inspected 
gauge copies since, judged from \Fig{fig:gh_vs_cp}, $\Ncp=40$ seems
to be on the safe side.

We conclude: the ghost propagator is systematically
dependent on the choice of Gribov copies, while the impact on the gluon
propagator is not resolvable within our statistics. However, there
are indications that the dependence on Gribov copies decreases with
increasing physical volume. This is also in agreement with the data
listed in the two lattice 
studies \cite{Bakeev:2003rr,Cucchieri:1997dx} of the $SU(2)$ ghost
propagator $G$, while it is not explicitly stated there. In fact, in
Ref.~\cite{Bakeev:2003rr} the ratio $G^{\fc}/G^{\bc}$ at $\beta=2.2$
on a $8^4$ lattice is larger than that on a $16^4$ lattice at the
same physical momentum.


\subsection{Systematic effects of lattice spacings and volumes}
\label{sec:sytematic}

We remind that in \Fig{fig:gh_and_gl_as_func_of_q} we have dropped all
data related to a $16^4$ lattice with one or more vanishing
momentum components $k_{\mu}$. According to \cite{Leinweber:1998uu}
this keeps finite volume effects for the gluon propagator under
control. It is quite natural to analyze here the different
systematic effects on the gluon and ghost propagators of
changing either the lattice spacing $a$ or the physical volume
$V$. However, due to the preselected set of momenta for
the ghost propagator and the values chosen for $\beta$, we can study
this here only in a limited way and in a region of intermediate
momenta. For the gluon propagator this has 
been done in more detail by other authors (see \eg~\cite{Bonnet:2001uh}).

\begin{figure}[tb]
  \hspace{-0.5cm}\includegraphics[width=8.5cm]{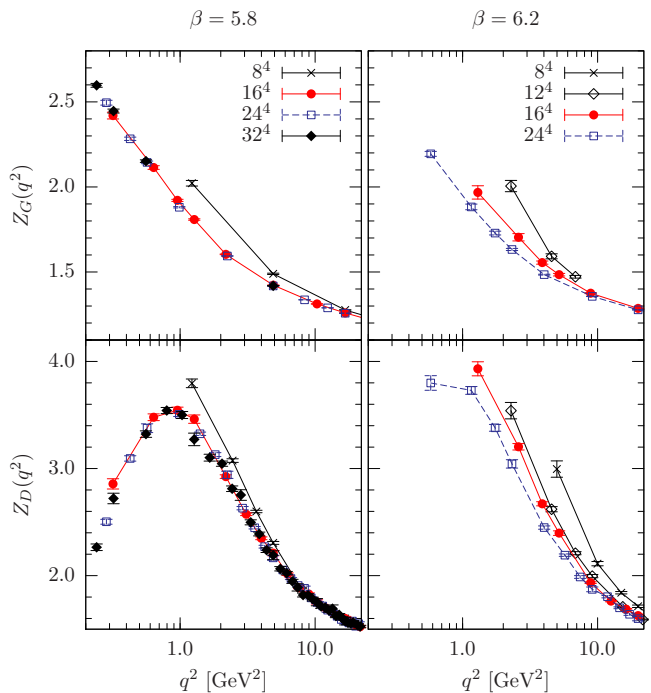}
  \caption{The ghost (upper panels) and gluon (lower panels)
  dressing functions are shown for different
    lattice sizes as functions of momentum $q^2$. The left panels show data at
    \mbox{$\beta=5.8$} and the right ones at \mbox{$\beta=6.2$}. Only
    data on \bc{} gauge copies are shown here. The lines are drawn to guide
    the eye.} 
  \label{fig:gh_gl_V}
\end{figure}

Keeping first the lattice spacing fixed we have found that both the ghost
and gluon dressing functions calculated at the same physical momentum $q^2$
decrease as the lattice size $L^4$ is increased. This is illustrated
for various momenta in \Fig{fig:gh_gl_V}. There both
dressings functions versus the physical 
momentum are shown for different lattice sizes at either $\beta=5.8$ or
$\beta=6.2$. Note, in this figure we have not dropped data with vanishing
momentum components $k_{\mu}$ to emphasize the influence of a finite
volume on those (low) momenta. We also show data from simulations
on a $8^4$ and $12^4$ lattice. One clearly sees that the lower the
momenta the larger the effect due to a finite volume. In comparison
with $\beta=5.8$ this is even more drastic at $\beta=6.2$. At this
$\beta$ the lattice spacing is about $a=0.068$~fm. Thus the largest
volume considered at $\beta=6.2$ is about (1.6~fm)$^4$, which is even
smaller than the physical volume of a $16^4$ lattice at $\beta=5.8$. 

Altogether we can state that for both dressing functions finite volume
effects are clearly visible at volumes smaller than (2.2~fm)$^4$,
which corresponds to a $16^4$ lattice at $\beta=5.8$. The effect grows
with decreasing momenta or decreasing lattice size (see the right panels in
\Fig{fig:gh_gl_V}). At larger volumes, however, the data for $q>1$~GeV
coincide within errors for the different lattice sizes (left
panels). Even for $q<1$~GeV we cannot resolve finite volume effects
for both dressing functions based on the data related to a $24^4$
and $32^4$ lattice. 

\begin{figure}[bt]
  \centering
  \mbox{\includegraphics[width=8.5cm]{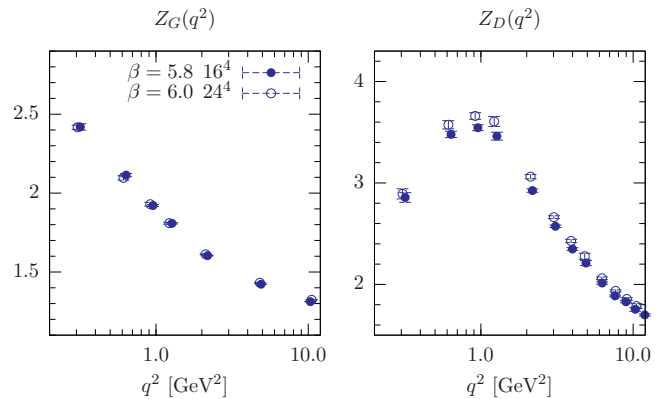}}
  \caption{The ghost (left) and the gluon (right) 
    dressing functions at (approximately) fixed physical volume
    \mbox{$V_1\approx$(2.2~fm)$^4$} are shown as functions of momentum~$q^2$.
    The data at $\beta=5.8$ (6.0) correspond to a
    lattice spacing of about a=0.136~fm (0.093~fm). Only
    data on \bc{} gauge copies are shown here.}
  \label{fig:gh_gL_a}
\end{figure}

\begin{figure*}
  \mbox{\includegraphics[width=8cm]{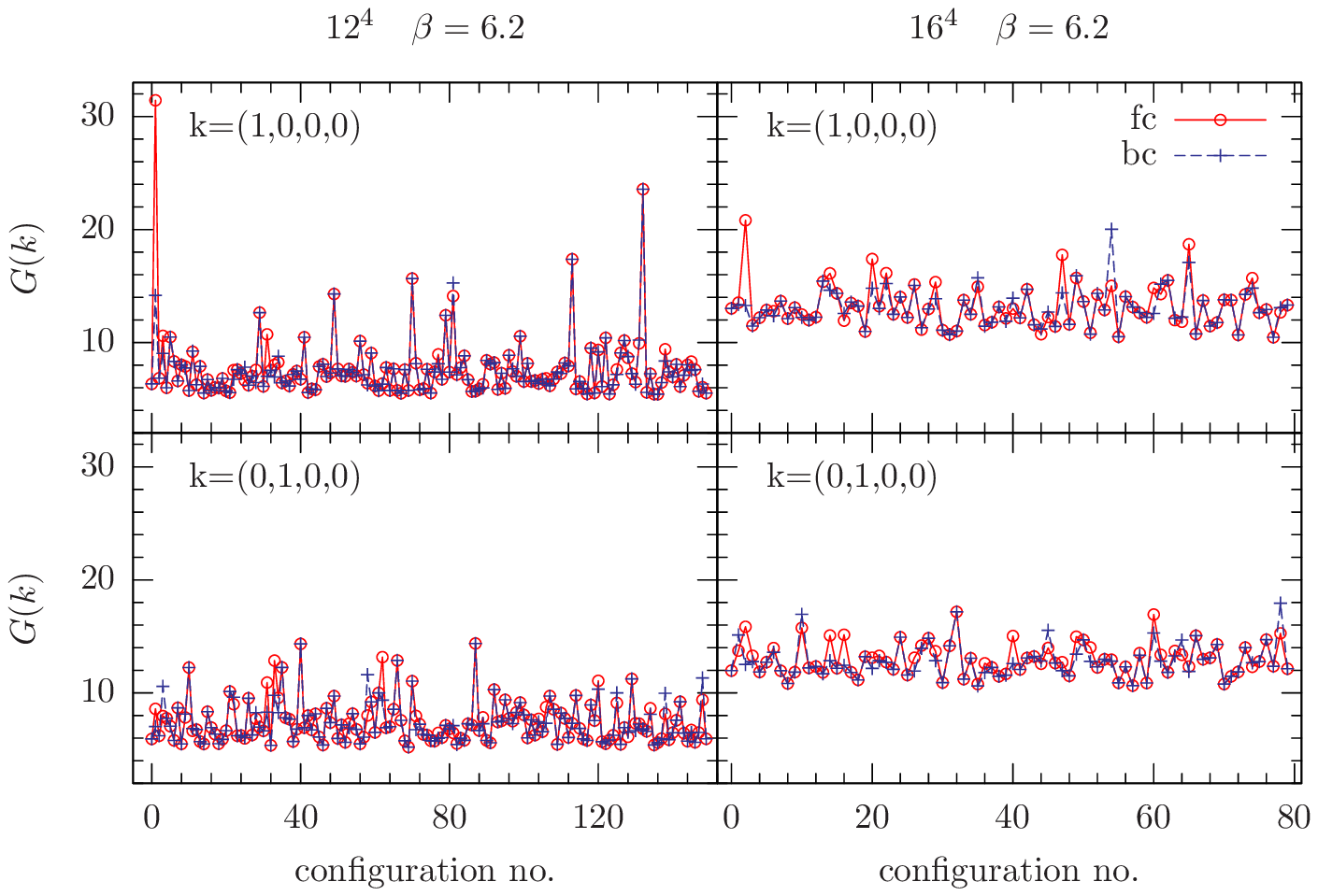}\quad
    \includegraphics[width=8cm]{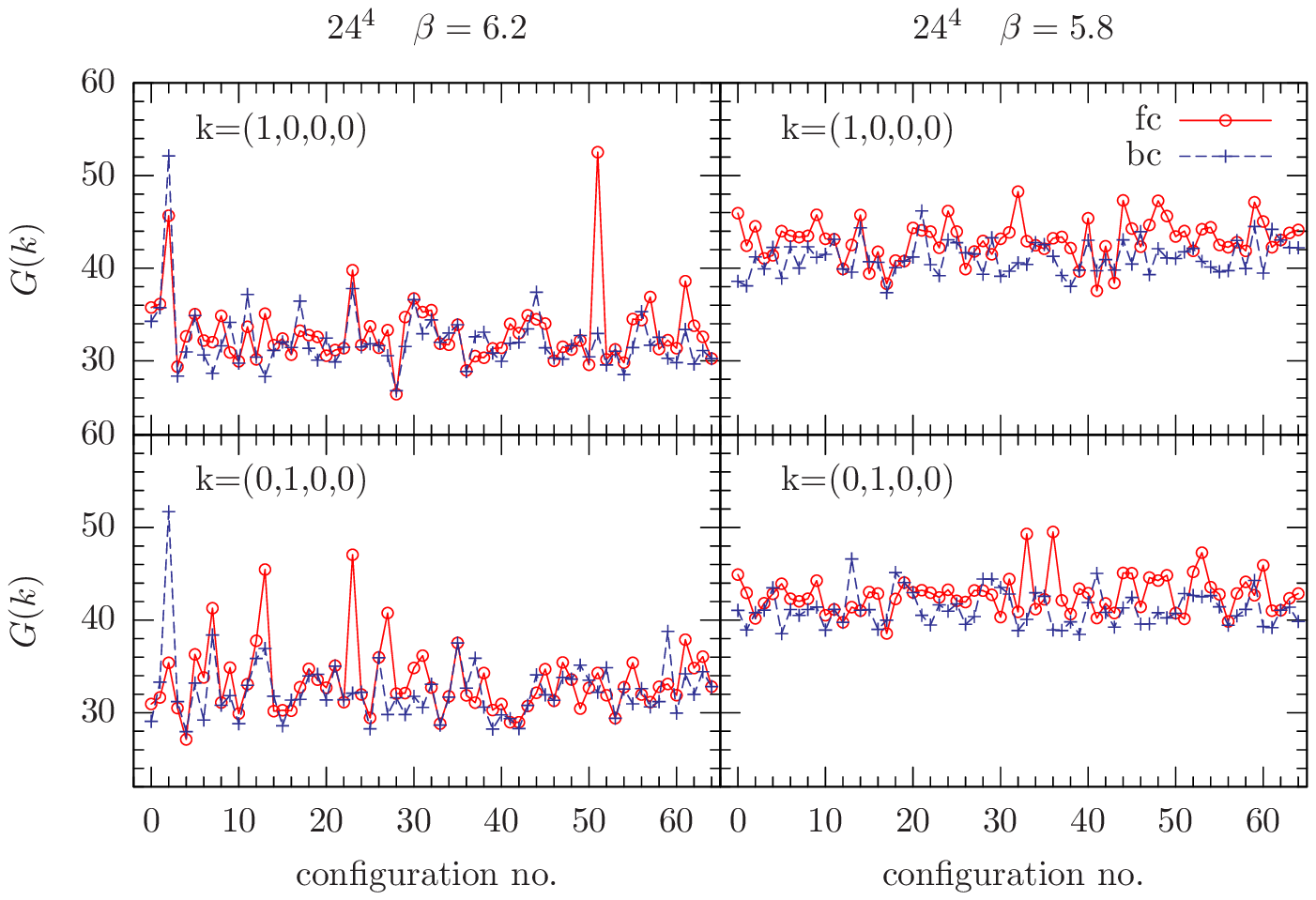}}
  \caption{The time histories of the ghost propagator calculated on
    first (\fc) and best (\bc) gauge copies on a $12^4$,
           $16^4$ and $24^4$ lattice at \mbox{$\beta=6.2$} and $24^4$
           lattice at \mbox{$\beta=5.8$}. From left to right the
           corresponding runs listed in \Tab{tab:statistic} are No.2,
           No.3, No.4/No.10 and No.1/No.8. The upper and lower panels
           show data for the lowest momentum realization
           \mbox{$k=(1,0,0,0)$} and \mbox{$k=(0,1,0,0)$}, respectively.}
  \label{fig:ghost_history}
\end{figure*}

Based on our chosen values for $\beta$ and the lattice sizes
we can select equal physical volumes only approximately. Hence also
the physical momenta 
are only approximately the same if the ghost and gluon dressing
functions are compared at different $\beta$, \eg at different lattice
spacings. Therefore, it is difficult to analyze
the systematic effect of changing~$a$ if for both dressing functions
small variations in~$q^2$ are hidden. Consequently, in
\Fig{fig:gh_gL_a} we show the data for the ghost and gluon dressing
functions approximately at the same physical volume
\mbox{$V\approx(2.2fm)^4$} for 
different $a$, albeit as functions of~$q^2$. This 
allows us to disentangle by eyes a change of the data due to
varying~$a$ from the natural dependence of the propagators on~$q^2$. 
Inspecting \Fig{fig:gh_gL_a} one concludes that the gluon dressing
function at the same physical momentum and volume increases with
decreasing lattice spacing. A similar effect (beyond error bars)
we cannot report for the ghost dressing function.

\section{The problem of exceptional configurations}
\label{sect:exceptional}

We turn now to a peculiarity of the ghost propagator at larger $\beta$
which has also been observed by some of us in an earlier $SU(2)$ study
\cite{Bakeev:2003rr}. While inspecting our data we found, though
rarely, that there are outliers in the Monte Carlo time histories of
the ghost propagator at lowest momentum. Those outliers are not
equally distributed around the average value, but are rather
significantly larger than this.

In \Fig{fig:ghost_history} we present time histories of
the ghost propagator $G(k)$ measured of \fc{} and \bc{} gauge copies
for two smallest momentum realizations \mbox{$k=(1,0,0,0)$} and
\mbox{$k=(0,1,0,0)$}, separately. From left to right the panels are
ordered in increasing order of the lattice sizes $12^4$, $16^4$ and
$24^4$ at $\beta=6.2$ and $24^4$ at $\beta=5.8$.

As can be seen from this figure in the majority extreme spikes are
reduced (or even not seen) when the ghost propagator could be measured
on a better gauge copy (\bc{}) for a particular
configuration. Furthermore, it is obvious that the
\emph{exceptionality} of a given gauge copy is exhibited not
simultaneously for all different realizations of the lowest
momentum. Consequently, to reduce the impact of such large values on the
average ghost propagator one should better average over all momentum
realizations giving rise to the same momentum $q$. This has been done
for the results shown in \Fig{fig:gh_vs_cp} and
\ref{fig:gh_and_gl_as_func_of_q} at least for the lowest
momentum at $\beta=6.2$. However, compared to the gluon propagator it
takes much more computing time to determine the ghost propagator for
all its different realizations of momentum $q(k)$.

In addition, we have tried to find a correlation of such outliers in the
history of the ghost propagator with other quantities measured
in our simulations. For example we have checked whether there is a
correlation between the values of the ghost propagator $G(k)$ as they
appear in \Fig{fig:ghost_history} 
with low-lying eigenvalues and
eigenvectors of the F-P operator. They are apparent in the
contribution of the lowest 10 F-P eigenmodes to the ghost
propagator at this particular $k$. This we shall present in a
subsequent publication \cite{Sternbeck:2005et} where we shall discuss
the spectral properties of the F-P operator and its relation to the
ghost propagator.

\section{The running coupling}
\label{sect:running_coupling}

We shall now focus on the running coupling $\alpha_s(q^2)$ as defined in
\Eq{eq:runcoupling} where $g^2_0/(4\pi) =
3/(2\pi\beta)$ for $SU(3)$. Given the raw data for the gluon and ghost
dressing functions on \bc{} gauge copies the average
$Z^2_{G}(q^2)Z_{D}(q^2)$ and its error have been estimated using the
\emph{bootstrap} method with drawing 500 random samples. Since the
ghost-ghost-gluon-vertex renormalization function $Z_1$ has been set to one,
there is an overall normalization factor which has been fixed by
fitting the data for $q^2>q^2_c$ to the well-known perturbative results of the
running coupling $\alpha_{\tt 2-loop}$ at 2-loop order (see also
\cite{Bloch:2003sk}). Defining \mbox{$x\equiv q^2/\Lambda^2_{\tt 2-loop}$},
the 2-loop running coupling is given by
\begin{equation}
  \label{eq:2loop}
  \alpha_{\tt 2-loop}(x) = \frac{4\pi}{\beta_0\ln x}
  \left\{1 - \frac{2\beta_1}{\beta^2_0}\frac{\ln(\ln x)}{\ln x} \right\}.
\end{equation}
The $\beta$-function coefficients are $\beta_0=11$ and $\beta_1=51$
for the $SU(3)$ gauge group and are independent of the renormalization
prescription. The value of $\Lambda_{\tt 2-loop}$ has been fixed by
the same fit. The lower bound $q^2_c$ has been chosen such that an
optimal value for $\chi^2/{\tt dof}$ has been achieved.

\begin{figure}[tb]
  \includegraphics[width=8.2cm]{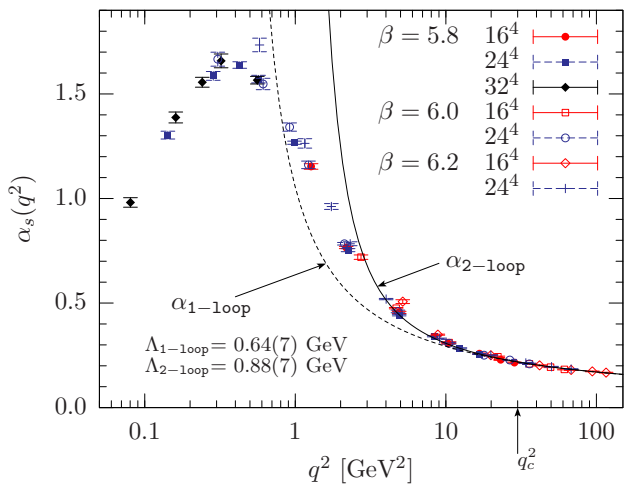}
  \caption{The running coupling $\alpha_s(q)$ as a function of
    momentum $q^2$ determined on \bc{} gauge copies.}
  \label{fig:run_coup}
\end{figure}

The results are shown in \Fig{fig:run_coup}. There also the \mbox{1-loop}
contribution is shown where we used the same lower
bound $q^2_c$. The best fit of the \mbox{2-loop} expression to the data gives
$\Lambda_{\tt 2-loop}$=0.88(7)~GeV ($\chi^2/{\tt dof}=0.96$), while
$\Lambda_{\tt 1-loop}$=0.64(7) is obtained ($\chi^2/{\tt dof}=1.05$)
using just the \mbox{1-loop} part. For $q^2_c$ we used $q^2_c=30$~GeV$^2$.
The value for $\Lambda_{\tt 2-loop}$ is similar within errors to the $SU(2)$
result given in~Ref.~\cite{Bloch:2003sk}.

Approaching the infrared limit in \Fig{fig:run_coup} one clearly sees a
running coupling $\alpha_s(q^2)$ increasing for
\mbox{$q^2>0.4$~GeV$^2$}. However, after passing a maximum at
\mbox{$q^2\approx0.4$~GeV$^2$} $\alpha_s(q^2)$ decreases again. Such
turnover is in agreement with DSE results obtained on a torus
\cite{Fischer:2002eq,Fischer:2002hn,Fischer:2005}. Therefore, one can
argue that this behavior is a finite lattice effect although we
cannot resolve a difference between the different lattice sizes used.
A similar infrared behavior for the running coupling has also been
observed in different lattice studies
\cite{Furui:2003jr,Furui:2004cx}. But opposed to \cite{Furui:2004cx}
the existence of a turnover is independent on the choice of Gribov
copy, since qualitatively, we have found the same behavior for
$\alpha_s(q^2)$ calculated on \fc{} gauge copies.

For completeness we mention that 
running couplings decreasing in the infrared have also been found 
in lattice studies of the \mbox{3-gluon}
vertex~\cite{Boucaud:1998bq,Boucaud:2002fx} and the quark-gluon
vertex~\cite{Skullerud:2002ge}.

Apart from the finite volume argument given above to explain such a
behavior, which prevents us from seeing the limit $\alpha_s(0) \ne 0 $ 
mentioned in the Introduction,
one could also put into question whether one can really set
$Z_1(q)=1$ at lower momenta. A recent investigation dedicated to the
ghost-ghost-gluon-vertex renormalization function $Z_1(q)$ for the
case of $SU(2)$ \cite{Cucchieri:2004sq} 
supports that $Z_1(q)\approx 1$ at least for $q>1$~GeV.

\section{Conclusions}
\label{sect:conclusions}

We have reported on a numerical study of the gluon and ghost
propagators in Landau gauge using several lattice sizes at
$\beta=5.8$, $6.0$ and $6.2$. Studying the dependence on the choice
of Gribov copies, it turns out that for the gluon propagator the
effect of Gribov copies stays inside numerical uncertainty, while the
impact on the ghost propagator increases as the momentum or $\beta$
is decreased. However, there are indications that the influence of
Gribov copies decreases as the physical volume is increased. This is
at least expected in the light of Ref.~\cite{Zwanziger:2003cf}. There
it is argued that 
in the continuum expectation values of correlation functions $\langle
A(x_1)\ldots A(x_n)\rangle$ over the fundamental modular region
$\Lambda$ are equal to those over the Gribov region $\Omega$, since
functional integrals are dominated by the common boundary of 
$\Lambda$ and $\Omega$. Thus Gribov copies inside $\Omega$ should not
affect expectation values in the continuum.

While the effect of the Gribov ambiguity on the ghost propagator
becomes smaller with increasing $\beta$, exceptionally large
values appear in the history of the ghost propagator in
agreement with what has been observed first in
reference \cite{Bakeev:2003rr}. These outliers we have not seen
simultaneously for
all lattice momenta $k$ realizing the same lowest momentum
$q(k)$. However, they are apparent in the contribution of the lowest
10 F-P eigenmodes to the ghost propagator at this particular $k$
\cite{Sternbeck:2005et}. Therefore it is good practice to measure the
ghost propagator for more than one $k$ with the same momentum $q(k)$,
in order to reduce the systematic errors coming from such exceptional
values.

We have studied the effects of finite volume on the one hand and of 
finite lattice spacing on the other. The first ones are found 
to be essential for volumes smaller than $(2.2~\mathrm{fm})^4$
at the same $\beta$ whereas the discretization effects at the 
same volume are modest. 
Our available data did not allow us to extend this 
analysis to physical momenta below 1 GeV, where the Gribov
ambiguity shows up and where a similar separation of finite 
volume and discretization effects would be desirable. 
However, we could observe from \Fig{fig:gh_and_gl_as_func_of_q}
that enlarging the volume by decreasing $\beta$ leads to a 
reduction of the systematic Gribov effect in the ghost propagator.

The dressing functions, $Z_{D}$ and $Z_{G}$, of the gluon and ghost
propagators have allowed us to estimate the behavior of a running coupling
$\alpha_s(q^2)$ in a momentum subtraction scheme. Going from larger
momenta $q^2$ to lower
ones $\alpha_s(q^2)$ is steadily increasing until
$q^2\approx0.4$~GeV$^2$. For $q^2<0.3$~GeV$^2$, however,
$\alpha_s(q^2)$ is decreasing. A decreasing running coupling at low
momenta is in qualitative agreement with recent DSE results obtained
on a torus~\cite{Fischer:2002eq,Fischer:2002hn,Fischer:2005}. Therefore one
might conclude that the decrease
is due to \emph{finite} lattice volumes we used. It makes it
questionable whether lattice simulations in near future can confirm
the predicted infrared behavior of the gluon and ghost dressing
functions with related exponents $\kappa_D = 2\kappa_G$.

\begin{appendix}
\section{Speeding up the inversion of the F-P operator}
\label{sect:pcg}

For the solution of the linear system $M\vec{\phi}=\vec{\psi}_c$
with symmetric matrix $M$, the conjugate gradient (CG) algorithm is
the method of choice. Its
convergence rate depends on the condition number, the ratio of largest
to lowest eigenvalue of $M$. When all $U_{x,\mu}=\identity$ obviously the
F-P operator is minus the Laplacian $\Delta$ with a diagonal color
substructure. Thus instead of solving
$M\vec{\phi}=\vec{\psi}_c$ one rather solves the
transformed system
\begin{displaymath}
  [M\Delta^{-1}]\,(\Delta\vec{\phi}\,)=\vec{\psi}_c
\end{displaymath}
In this way the condition number is reduced, however,
the price to pay is one extra matrix multiplication by
$\Delta^{-1}$ per iteration cycle. In terms of CPU time this should be
more than compensated by the reduction of iterations.

The pre-conditioned CG algorithm (PCG) can be described as follows:
\begin{eqnarray*}
  \textit{initialize:}\hfill \\
  \vec{r}^{\,(0)} &=& \vec{\psi} - M\vec{\phi}^{\,(0)},\quad
  \vec{p}^{\,(0)} = \Delta^{-1}\,\vec{r}^{\,(0)},\\
  \gamma^{(0)} &=& ( \vec{p}^{\,(0)},\vec{r}^{\,(0)})\\*[0.1cm]
  \textit{start do loop:} && k=0,1,\ldots\\
  \vec{z}^{\,(k)} &=&  M\vec{p}^{\,(k)},\quad
  \alpha^{(k)}=\gamma^{\,(k)}/(\vec{z}^{\,(k)},\vec{p}^{\,(k)})\\
  \vec{\phi}^{\,(k+1)} &=& \vec{\phi}^{\,(k)} +
  \alpha^{(k)}\vec{p}^{\,(k)}\\
  \vec{r}^{\,(k+1)} &=& \vec{r}^{\,(k)} -
  \alpha^{(k)}\vec{z}^{\,(k)}\\
  \vec{z}^{\,(k+1)} &=& \Delta^{-1}\,\vec{r}^{\,(k+1)}\\
  \gamma^{(k+1)} &=& ( \vec{z}^{\,(k+1)}, \vec{r}^{\,(k+1)})\\
  &\textit{if}&(\gamma^{(k+1)}<\varepsilon)\quad\textit{exit do loop}\\
  \vec{p}^{\,(k+1)} &=& \vec{z}^{\,(k+1)} +
  \frac{\gamma^{(k+1)}}{\gamma^{(k)}}\vec{p}^{\,(k)}\\
  \textit{end do loop}
\end{eqnarray*}
Here $(\cdot,\cdot)$ denotes the scalar product.

To perform the additional matrix multiplication with $\Delta^{-1}$ we
used two fast Fourier transformations $\mathcal{F}$, due to
\mbox{$(-\Delta)^{-1}=\mathcal{F}^{-1}\,q^{-2}(k)\,\mathcal{F}$}.
The performance we achieved is presented in \Tab{tab:speedup}. We conclude
that on larger lattice sizes the reduction of
iterations is about 70-75\%, while the resulting reduction of CPU time depends
on the lattice size. This is because we are using the fast Fourier
transformations in a parallel CPU environment. If the ratio
of used processors to the lattice size is small (see \eg
the data for $32^4$ lattice at this table),
almost the same reductions of CPU time as for the number of iterations
is achieved.

Further improvement may be achieved by using the multigrid Poisson solver to
solve \mbox{$\Delta\vec{z}^{\,(k)}=\vec{r}^{\,(k)}$}. This method
is supposed to perform better on parallel machines. Perhaps a further
improvement is possible by using as pre-conditioning matrix
\mbox{$\widetilde{M}^{-1}=-\Delta^{-1}-\Delta^{-1}M_1\Delta^{-1} + \ldots$}
which is an approximation of the F-P operator \mbox{$M=-\Delta+M_1$} to a
given order \cite{Zwanziger:1993dh} (see also
\cite{Furui:2003jr}). However, the larger the
order, the more matrix multiplications per iteration cycle are
required. This may reduce the overall performance.
We have not checked so far which is the optimal order.

\begin{table}[htb]
  \centering
  \newcommand{\mc}[3]{\multicolumn{#1}{#2}{#3}}
  \begin{tabular}{c@{\quad}cc@{\quad}cc@{\quad}cc}
    \hline\hline
            & \mc{2}{l}{CG}  & \mc{2}{l}{PCG} & \mc{2}{l}{speed up}\\
    lattice & iter & CPU[sec] & iter & CPU[sec] & iter & CPU[sec]\\
    \hline
    $8^4$   &   1400 &  3.7   & 570  & 2.4  & 60\%  & 35\% \\
    $16^4$  &   3900 & 240    & 1050 & 130  & 73\%  & 46\% \\
    $32^4$  &   9900 & 13400  & 2250 & 3900 & 77\%  & 71\% \\
    \hline\hline
  \end{tabular}
  \caption{The average number of iterations and CPU time per
            processor (PE) using the CG and PCG algorithm to invert
            the F-P operator are given. All inversions have been
            performed at $\beta=5.8$ with source
            $\delta^{bc} \exp (2\pi i\,k\!\cdot\! y)$ where
            $k=(1,0,0,0)$. To compare the different
            lattice sizes 4 PEs have always been used.}
  \label{tab:speedup}
\end{table}

\end{appendix}

\section*{ACKNOWLEDGMENTS}

All simulations have been done on the IBM pSeries 690 at HLRN.  We
thank R.~Alkofer for discussions and H.~St\"uben for
contributing parts of the program code. We are indebted to
C.~Fischer for communicating us his DSE results
\cite{Fischer:2005} prior to publication. This work has been supported
by the DFG under contract FOR~465.  A.~Sternbeck acknowledges support
of the DFG-funded graduate school GK~271.


\bibliographystyle{apsrev}

\begin{thebibliography}{36}
\expandafter\ifx\csname natexlab\endcsname\relax\def\natexlab#1{#1}\fi
\expandafter\ifx\csname bibnamefont\endcsname\relax
  \def\bibnamefont#1{#1}\fi
\expandafter\ifx\csname bibfnamefont\endcsname\relax
  \def\bibfnamefont#1{#1}\fi
\expandafter\ifx\csname citenamefont\endcsname\relax
  \def\citenamefont#1{#1}\fi
\expandafter\ifx\csname url\endcsname\relax
  \def\url#1{\texttt{#1}}\fi
\expandafter\ifx\csname urlprefix\endcsname\relax\def\urlprefix{URL }\fi
\providecommand{\bibinfo}[2]{#2}
\providecommand{\eprint}[2][]{\url{#2}}

\bibitem[{\citenamefont{Alkofer and von Smekal}(2001)}]{Alkofer:2000wg}
\bibinfo{author}{\bibfnamefont{R.}~\bibnamefont{Alkofer}} \bibnamefont{and}
  \bibinfo{author}{\bibfnamefont{L.}~\bibnamefont{von Smekal}},
  \bibinfo{journal}{Phys. Rept.} \textbf{\bibinfo{volume}{353}},
  \bibinfo{pages}{281} (\bibinfo{year}{2001}), \eprint{hep-ph/0007355}.

\bibitem[{\citenamefont{Fischer and Alkofer}(2003)}]{Fischer:2003rp}
\bibinfo{author}{\bibfnamefont{C.~S.} \bibnamefont{Fischer}} \bibnamefont{and}
  \bibinfo{author}{\bibfnamefont{R.}~\bibnamefont{Alkofer}},
  \bibinfo{journal}{Phys. Rev.} \textbf{\bibinfo{volume}{D67}},
  \bibinfo{pages}{094020} (\bibinfo{year}{2003}), \eprint{hep-ph/0301094}.

\bibitem[{\citenamefont{Lerche and von Smekal}(2002)}]{Lerche:2002ep}
\bibinfo{author}{\bibfnamefont{C.}~\bibnamefont{Lerche}} \bibnamefont{and}
  \bibinfo{author}{\bibfnamefont{L.}~\bibnamefont{von Smekal}},
  \bibinfo{journal}{Phys. Rev.} \textbf{\bibinfo{volume}{D65}},
  \bibinfo{pages}{125006} (\bibinfo{year}{2002}), \eprint{hep-ph/0202194}.

\bibitem[{\citenamefont{Zwanziger}(2002)}]{Zwanziger:2001kw}
\bibinfo{author}{\bibfnamefont{D.}~\bibnamefont{Zwanziger}},
  \bibinfo{journal}{Phys. Rev.} \textbf{\bibinfo{volume}{D65}},
  \bibinfo{pages}{094039} (\bibinfo{year}{2002}), \eprint{hep-th/0109224}.

\bibitem[{\citenamefont{Zwanziger}(2004)}]{Zwanziger:2003cf}
\bibinfo{author}{\bibfnamefont{D.}~\bibnamefont{Zwanziger}},
  \bibinfo{journal}{Phys. Rev.} \textbf{\bibinfo{volume}{D69}},
  \bibinfo{pages}{016002} (\bibinfo{year}{2004}), \eprint{hep-ph/0303028}.

\bibitem[{\citenamefont{Zwanziger}(1994)}]{Zwanziger:1993dh}
\bibinfo{author}{\bibfnamefont{D.}~\bibnamefont{Zwanziger}},
  \bibinfo{journal}{Nucl. Phys.} \textbf{\bibinfo{volume}{B412}},
  \bibinfo{pages}{657} (\bibinfo{year}{1994}).

\bibitem[{\citenamefont{Gribov}(1978)}]{Gribov:1977wm}
\bibinfo{author}{\bibfnamefont{V.~N.} \bibnamefont{Gribov}},
  \bibinfo{journal}{Nucl. Phys.} \textbf{\bibinfo{volume}{B139}},
  \bibinfo{pages}{1} (\bibinfo{year}{1978}).

\bibitem[{\citenamefont{Kugo and Ojima}(1979)}]{Kugo:1979gm}
\bibinfo{author}{\bibfnamefont{T.}~\bibnamefont{Kugo}} \bibnamefont{and}
  \bibinfo{author}{\bibfnamefont{I.}~\bibnamefont{Ojima}},
  \bibinfo{journal}{Prog. Theor. Phys. Suppl.} \textbf{\bibinfo{volume}{66}},
  \bibinfo{pages}{1} (\bibinfo{year}{1979}).

\bibitem[{\citenamefont{von Smekal et~al.}(1997)\citenamefont{von Smekal,
  Alkofer, and Hauck}}]{vonSmekal:1997is}
\bibinfo{author}{\bibfnamefont{L.}~\bibnamefont{von Smekal}},
  \bibinfo{author}{\bibfnamefont{A.}~\bibnamefont{Hauck}}, \bibnamefont{and}
  \bibinfo{author}{\bibfnamefont{R.}~\bibnamefont{Alkofer}}, 
  \bibinfo{journal}{Phys. Rev. Lett.} \textbf{\bibinfo{volume}{79}},
  \bibinfo{pages}{3591} (\bibinfo{year}{1997}), \eprint{hep-ph/9705242}.

\bibitem[{\citenamefont{von Smekal et~al.}(1998)\citenamefont{von Smekal,
  Hauck, and Alkofer}}]{vonSmekal:1998}
\bibinfo{author}{\bibfnamefont{L.}~\bibnamefont{von Smekal}},
  \bibinfo{author}{\bibfnamefont{A.}~\bibnamefont{Hauck}}, \bibnamefont{and}
  \bibinfo{author}{\bibfnamefont{R.}~\bibnamefont{Alkofer}},
  \bibinfo{journal}{Ann. Phys.} \textbf{\bibinfo{volume}{267}},
  \bibinfo{pages}{1} (\bibinfo{year}{1998}), \eprint{hep-ph/9707327}.

\bibitem[{\citenamefont{Bloch et~al.}(2004)\citenamefont{Bloch, Cucchieri,
  Langfeld, and Mendes}}]{Bloch:2003sk}
\bibinfo{author}{\bibfnamefont{J.~C.~R.} \bibnamefont{Bloch}},
  \bibinfo{author}{\bibfnamefont{A.}~\bibnamefont{Cucchieri}},
  \bibinfo{author}{\bibfnamefont{K.}~\bibnamefont{Langfeld}}, \bibnamefont{and}
  \bibinfo{author}{\bibfnamefont{T.}~\bibnamefont{Mendes}},
  \bibinfo{journal}{Nucl. Phys.} \textbf{\bibinfo{volume}{B687}},
  \bibinfo{pages}{76} (\bibinfo{year}{2004}), \eprint{hep-lat/0312036}.

\bibitem[{\citenamefont{Taylor}(1971)}]{Taylor:1971ff}
\bibinfo{author}{\bibfnamefont{J.~C.} \bibnamefont{Taylor}},
  \bibinfo{journal}{Nucl. Phys.} \textbf{\bibinfo{volume}{B33}},
  \bibinfo{pages}{436} (\bibinfo{year}{1971}).

\bibitem[{\citenamefont{Cucchieri et~al.}(2004)\citenamefont{Cucchieri, Mendes,
  and Mihara}}]{Cucchieri:2004sq}
\bibinfo{author}{\bibfnamefont{A.}~\bibnamefont{Cucchieri}},
  \bibinfo{author}{\bibfnamefont{T.}~\bibnamefont{Mendes}}, \bibnamefont{and}
  \bibinfo{author}{\bibfnamefont{A.}~\bibnamefont{Mihara}},
  \bibinfo{journal}{JHEP} \textbf{\bibinfo{volume}{12}}, \bibinfo{pages}{012}
  (\bibinfo{year}{2004}), \eprint{hep-lat/0408034}.

\bibitem[{\citenamefont{Langfeld et~al.}(2002)\citenamefont{Langfeld,
  Reinhardt, and Gattnar}}]{Langfeld:2001cz}
\bibinfo{author}{\bibfnamefont{K.}~\bibnamefont{Langfeld}},
  \bibinfo{author}{\bibfnamefont{H.}~\bibnamefont{Reinhardt}},
  \bibnamefont{and} \bibinfo{author}{\bibfnamefont{J.}~\bibnamefont{Gattnar}},
  \bibinfo{journal}{Nucl. Phys.} \textbf{\bibinfo{volume}{B621}},
  \bibinfo{pages}{131} (\bibinfo{year}{2002}), \eprint{hep-ph/0107141}.

\bibitem[{\citenamefont{Gattnar et~al.}(2004)\citenamefont{Gattnar, Langfeld,
  and Reinhardt}}]{Gattnar:2004bf}
\bibinfo{author}{\bibfnamefont{J.}~\bibnamefont{Gattnar}},
  \bibinfo{author}{\bibfnamefont{K.}~\bibnamefont{Langfeld}}, \bibnamefont{and}
  \bibinfo{author}{\bibfnamefont{H.}~\bibnamefont{Reinhardt}},
  \bibinfo{journal}{Phys. Rev. Lett.} \textbf{\bibinfo{volume}{93}},
  \bibinfo{pages}{061601} (\bibinfo{year}{2004}), \eprint{hep-lat/0403011}.

\bibitem[{\citenamefont{Bloch et~al.}(2003)\citenamefont{Bloch, Cucchieri,
  Langfeld, and Mendes}}]{Bloch:2002we}
\bibinfo{author}{\bibfnamefont{J.~C.~R.} \bibnamefont{Bloch}},
  \bibinfo{author}{\bibfnamefont{A.}~\bibnamefont{Cucchieri}},
  \bibinfo{author}{\bibfnamefont{K.}~\bibnamefont{Langfeld}}, \bibnamefont{and}
  \bibinfo{author}{\bibfnamefont{T.}~\bibnamefont{Mendes}},
  \bibinfo{journal}{Nucl. Phys. Proc. Suppl.} \textbf{\bibinfo{volume}{119}},
  \bibinfo{pages}{736} (\bibinfo{year}{2003}), \eprint{hep-lat/0209040}.

\bibitem[{\citenamefont{Oliveira and Silva}(2005)}]{Oliveira:2004gy}
\bibinfo{author}{\bibfnamefont{O.}~\bibnamefont{Oliveira}} \bibnamefont{and}
  \bibinfo{author}{\bibfnamefont{P.~J.} \bibnamefont{Silva}},
  \bibinfo{journal}{AIP Conf. Proc.} \textbf{\bibinfo{volume}{756}},
  \bibinfo{pages}{290} (\bibinfo{year}{2005}), \eprint{hep-lat/0410048}.

\bibitem[{\citenamefont{Sternbeck
  et~al.}(2005{\natexlab{a}})\citenamefont{Sternbeck, Ilgenfritz,
  M{\"u}ller-Preussker, and Schiller}}]{Sternbeck:2004xr}
\bibinfo{author}{\bibfnamefont{A.}~\bibnamefont{Sternbeck}},
  \bibinfo{author}{\bibfnamefont{E.-M.} \bibnamefont{Ilgenfritz}},
  \bibinfo{author}{\bibfnamefont{M.}~\bibnamefont{M{\"u}ller-Preussker}},
  \bibnamefont{and} \bibinfo{author}{\bibfnamefont{A.}~\bibnamefont{Schiller}},
  \bibinfo{journal}{Nucl. Phys. Proc. Suppl.} \textbf{\bibinfo{volume}{140}},
  \bibinfo{pages}{653} (\bibinfo{year}{2005}{\natexlab{a}}),
  \eprint{hep-lat/0409125}.

\bibitem[{\citenamefont{Sternbeck
  et~al.}(2005{\natexlab{b}})\citenamefont{Sternbeck, Ilgenfritz,
  M{\"u}ller-Preussker, and Schiller}}]{Sternbeck:2004qk}
\bibinfo{author}{\bibfnamefont{A.}~\bibnamefont{Sternbeck}},
  \bibinfo{author}{\bibfnamefont{E.-M.} \bibnamefont{Ilgenfritz}},
  \bibinfo{author}{\bibfnamefont{M.}~\bibnamefont{M{\"u}ller-Preussker}},
  \bibnamefont{and} \bibinfo{author}{\bibfnamefont{A.}~\bibnamefont{Schiller}},
  \bibinfo{journal}{AIP Conf. Proc.} \textbf{\bibinfo{volume}{756}},
  \bibinfo{pages}{284} (\bibinfo{year}{2005}{\natexlab{b}}),
  \eprint{hep-lat/0412011}.

\bibitem[{\citenamefont{Furui and Nakajima}(2004{\natexlab{a}})}]{Furui:2004cx}
\bibinfo{author}{\bibfnamefont{S.}~\bibnamefont{Furui}} \bibnamefont{and}
  \bibinfo{author}{\bibfnamefont{H.}~\bibnamefont{Nakajima}},
  \bibinfo{journal}{Phys. Rev.} \textbf{\bibinfo{volume}{D70}},
  \bibinfo{pages}{094504} (\bibinfo{year}{2004}{\natexlab{a}}),
  \eprint{hep-lat/0403021}.

\bibitem[{\citenamefont{Furui and Nakajima}(2004{\natexlab{b}})}]{Furui:2003jr}
\bibinfo{author}{\bibfnamefont{S.}~\bibnamefont{Furui}} \bibnamefont{and}
  \bibinfo{author}{\bibfnamefont{H.}~\bibnamefont{Nakajima}},
  \bibinfo{journal}{Phys. Rev.} \textbf{\bibinfo{volume}{D69}},
  \bibinfo{pages}{074505} (\bibinfo{year}{2004}{\natexlab{b}}),
  \eprint{hep-lat/0305010}.

\bibitem[{\citenamefont{Fischer et~al.}(2005)\citenamefont{Fischer, Gr{\"u}ter,
  and Alkofer}}]{Fischer:2005}
\bibinfo{author}{\bibfnamefont{C.~S.} \bibnamefont{Fischer}},
  \bibinfo{author}{\bibfnamefont{B.}~\bibnamefont{Gr{\"u}ter}},
  \bibnamefont{and} \bibinfo{author}{\bibfnamefont{R.}~\bibnamefont{Alkofer}}
  (\bibinfo{year}{2005}), \eprint{hep-ph/0506053}.

\bibitem[{\citenamefont{Fischer et~al.}(2002)\citenamefont{Fischer, Alkofer,
  and Reinhardt}}]{Fischer:2002eq}
\bibinfo{author}{\bibfnamefont{C.~S.} \bibnamefont{Fischer}},
  \bibinfo{author}{\bibfnamefont{R.}~\bibnamefont{Alkofer}}, \bibnamefont{and}
  \bibinfo{author}{\bibfnamefont{H.}~\bibnamefont{Reinhardt}},
  \bibinfo{journal}{Phys. Rev.} \textbf{\bibinfo{volume}{D65}},
  \bibinfo{pages}{094008} (\bibinfo{year}{2002}), \eprint{hep-ph/0202195}.

\bibitem[{\citenamefont{Fischer and Alkofer}(2002)}]{Fischer:2002hn}
\bibinfo{author}{\bibfnamefont{C.~S.} \bibnamefont{Fischer}} \bibnamefont{and}
  \bibinfo{author}{\bibfnamefont{R.}~\bibnamefont{Alkofer}},
  \bibinfo{journal}{Phys. Lett.} \textbf{\bibinfo{volume}{B536}},
  \bibinfo{pages}{177} (\bibinfo{year}{2002}), \eprint{hep-ph/0202202}.

\bibitem[{\citenamefont{Cucchieri}(1997)}]{Cucchieri:1997dx}
\bibinfo{author}{\bibfnamefont{A.}~\bibnamefont{Cucchieri}},
  \bibinfo{journal}{Nucl. Phys.} \textbf{\bibinfo{volume}{B508}},
  \bibinfo{pages}{353} (\bibinfo{year}{1997}), \eprint{hep-lat/9705005}.

\bibitem[{\citenamefont{Bakeev et~al.}(2004)\citenamefont{Bakeev, Ilgenfritz,
  Mitrjushkin, and M{\"u}ller-Preussker}}]{Bakeev:2003rr}
\bibinfo{author}{\bibfnamefont{T.~D.} \bibnamefont{Bakeev}},
  \bibinfo{author}{\bibfnamefont{E.-M.} \bibnamefont{Ilgenfritz}},
  \bibinfo{author}{\bibfnamefont{M.}~\bibnamefont{M{\"u}ller-Preussker}},
  \bibnamefont{and}
  \bibinfo{author}{\bibfnamefont{V.~K.} \bibnamefont{Mitrjushkin}},
  \bibinfo{journal}{Phys. Rev.} \textbf{\bibinfo{volume}{D69}},
  \bibinfo{pages}{074507} (\bibinfo{year}{2004}), \eprint{hep-lat/0311041}.

\bibitem[{\citenamefont{Nakajima and Furui}(2004)}]{Nakajima:2003my}
\bibinfo{author}{\bibfnamefont{H.}~\bibnamefont{Nakajima}} \bibnamefont{and}
  \bibinfo{author}{\bibfnamefont{S.}~\bibnamefont{Furui}},
  \bibinfo{journal}{Nucl. Phys. Proc. Suppl.} \textbf{\bibinfo{volume}{129}},
  \bibinfo{pages}{730} (\bibinfo{year}{2004}), \eprint{hep-lat/0309165}.

\bibitem[{\citenamefont{Silva and Oliveira}(2004)}]{Silva:2004bv}
\bibinfo{author}{\bibfnamefont{P.~J.} \bibnamefont{Silva}} \bibnamefont{and}
  \bibinfo{author}{\bibfnamefont{O.}~\bibnamefont{Oliveira}},
  \bibinfo{journal}{Nucl. Phys.} \textbf{\bibinfo{volume}{B690}},
  \bibinfo{pages}{177} (\bibinfo{year}{2004}), \eprint{hep-lat/0403026}.

\bibitem[{\citenamefont{Leinweber et~al.}(1999)\citenamefont{Leinweber,
  Skullerud, Williams, and Parrinello}}]{Leinweber:1998uu}
\bibinfo{author}{\bibfnamefont{D.~B.} \bibnamefont{Leinweber}},
  \bibinfo{author}{\bibfnamefont{J.~I.} \bibnamefont{Skullerud}},
  \bibinfo{author}{\bibfnamefont{A.~G.} \bibnamefont{Williams}},
  \bibnamefont{and}
  \bibinfo{author}{\bibfnamefont{C.}~\bibnamefont{Parrinello}},
  (\bibinfo{collaboration}{UKQCD}), \bibinfo{journal}{Phys. Rev.}
  \textbf{\bibinfo{volume}{D60}}, \bibinfo{pages}{094507}
  (\bibinfo{year}{1999}), \eprint{hep-lat/9811027}.

\bibitem[{\citenamefont{Suman and Schilling}(1996)}]{Suman:1995zg}
\bibinfo{author}{\bibfnamefont{H.}~\bibnamefont{Suman}} \bibnamefont{and}
  \bibinfo{author}{\bibfnamefont{K.}~\bibnamefont{Schilling}},
  \bibinfo{journal}{Phys. Lett.} \textbf{\bibinfo{volume}{B373}},
  \bibinfo{pages}{314} (\bibinfo{year}{1996}), \eprint{hep-lat/9512003}.

\bibitem[{\citenamefont{Necco and Sommer}(2002)}]{Necco:2001xg}
\bibinfo{author}{\bibfnamefont{S.}~\bibnamefont{Necco}} \bibnamefont{and}
  \bibinfo{author}{\bibfnamefont{R.}~\bibnamefont{Sommer}},
  \bibinfo{journal}{Nucl. Phys.} \textbf{\bibinfo{volume}{B622}},
  \bibinfo{pages}{328} (\bibinfo{year}{2002}), \eprint{hep-lat/0108008}.

\bibitem[{\citenamefont{Bonnet et~al.}(2001)\citenamefont{Bonnet, Bowman,
  Leinweber, Williams, and Zanotti}}]{Bonnet:2001uh}
\bibinfo{author}{\bibfnamefont{F.~D.~R.} \bibnamefont{Bonnet}},
  \bibinfo{author}{\bibfnamefont{P.~O.} \bibnamefont{Bowman}},
  \bibinfo{author}{\bibfnamefont{D.~B.} \bibnamefont{Leinweber}},
  \bibinfo{author}{\bibfnamefont{A.~G.} \bibnamefont{Williams}},
  \bibnamefont{and} \bibinfo{author}{\bibfnamefont{J.~M.}
  \bibnamefont{Zanotti}}, \bibinfo{journal}{Phys. Rev.}
  \textbf{\bibinfo{volume}{D64}}, \bibinfo{pages}{034501}
  (\bibinfo{year}{2001}), \eprint{hep-lat/0101013}.

\bibitem[{\citenamefont{Sternbeck
  et~al.}(2005{\natexlab{c}})\citenamefont{Sternbeck, Ilgenfritz,
  M{\"u}ller-Preussker, and Schiller}}]{Sternbeck:2005et}
\bibinfo{author}{\bibfnamefont{A.}~\bibnamefont{Sternbeck}},
  \bibinfo{author}{\bibfnamefont{E.-M.} \bibnamefont{Ilgenfritz}},
  \bibinfo{author}{\bibfnamefont{M.}~\bibnamefont{M{\"u}ller-Preussker}},
  \bibnamefont{and} \bibinfo{author}{\bibfnamefont{A.}~\bibnamefont{Schiller}},
  \bibinfo{journal}{in progress}  (\bibinfo{year}{2005}{\natexlab{c}}).

\bibitem[{\citenamefont{Boucaud et~al.}(1998)\citenamefont{Boucaud, Leroy,
  Micheli, Pene, and Roiesnel}}]{Boucaud:1998bq}
\bibinfo{author}{\bibfnamefont{P.}~\bibnamefont{Boucaud}},
  \bibinfo{author}{\bibfnamefont{J.~P.} \bibnamefont{Leroy}},
  \bibinfo{author}{\bibfnamefont{J.}~\bibnamefont{Micheli}},
  \bibinfo{author}{\bibfnamefont{O.}~\bibnamefont{Pene}}, \bibnamefont{and}
  \bibinfo{author}{\bibfnamefont{C.}~\bibnamefont{Roiesnel}},
  \bibinfo{journal}{JHEP} \textbf{\bibinfo{volume}{10}}, \bibinfo{pages}{017}
  (\bibinfo{year}{1998}), \eprint{hep-ph/9810322}.

\bibitem[{\citenamefont{Boucaud et~al.}(2003)}]{Boucaud:2002fx}
\bibinfo{author}{\bibfnamefont{P.}~\bibnamefont{Boucaud}} \bibnamefont{et~al.},
  \bibinfo{journal}{JHEP} \textbf{\bibinfo{volume}{04}}, \bibinfo{pages}{005}
  (\bibinfo{year}{2003}), \eprint{hep-ph/0212192}.

\bibitem[{\citenamefont{Skullerud and Kizilersu}(2002)}]{Skullerud:2002ge}
\bibinfo{author}{\bibfnamefont{J.}~\bibnamefont{Skullerud}} \bibnamefont{and}
  \bibinfo{author}{\bibfnamefont{A.}~\bibnamefont{Kizilersu}},
  \bibinfo{journal}{JHEP} \textbf{\bibinfo{volume}{09}}, \bibinfo{pages}{013}
  (\bibinfo{year}{2002}), \eprint{hep-ph/0205318}.

\end{thebibliography}

\end{document}